\documentclass[10pt,conference]{IEEEtran}

\usepackage{srcltx}
\usepackage[psamsfonts]{amsfonts}
\usepackage{amsmath}
\usepackage{amssymb}
\usepackage{amsfonts}
\usepackage{graphics}
\usepackage{psfrag}
\usepackage{epsfig}
\usepackage{array}
\usepackage{algorithm}
\usepackage{algorithmic}
\usepackage{pifont}
\usepackage{srcltx}
\usepackage[psamsfonts]{amsfonts}
\usepackage{makeidx}  
\usepackage{bbm}
\usepackage{hhline}
\usepackage{eufrak}
\usepackage{yfonts}
\usepackage{color}
\usepackage{url}
\usepackage{dsfont}
\usepackage{caption}
\usepackage{subcaption}
\usepackage[square, comma, sort&compress, numbers]{natbib}

\interdisplaylinepenalty=2500 \algsetup{indent=2em}
\title{Missing Spectrum-Data Recovery in Cognitive Radio Networks Using Piecewise Constant Nonnegative Matrix Factorization}
\author{Alireza Zaeemzadeh, Mohsen Joneidi, Behzad Shahrasbi, and Nazanin Rahnavard\\
School of Electrical Engineering and Computer Science \\
University of Central Florida \\
Emails: {\{zaeemzadeh, joneidi, behzad.shahrasbi\}@knights.ucf.edu, and nazanin@eecs.ucf.edu}}
\begin{document}
\renewcommand{\textfraction}{0}
\maketitle

\begin{abstract}
In this paper, we propose a missing spectrum data recovery technique for cognitive radio (CR) networks using Nonnegative Matrix Factorization (NMF). It is shown that the spectrum measurements collected from secondary users (SUs) can be factorized as product of a channel gain matrix times an activation matrix. Then, an NMF method with piecewise constant activation coefficients is introduced to analyze the measurements and estimate the missing spectrum data. The proposed optimization problem is solved by a Majorization-Minimization technique. The numerical simulation verifies that the proposed technique is able to accurately estimate the missing spectrum data in the presence of noise and fading.\par
\end{abstract}
\section{Introduction}\label{intro}
Recent advances in wireless communications and microelectronic devices are leading the trend of research toward cognitive radios (CRs) \cite{Zhao07asurvey}. The main feature of CRs is the opportunistic usage of spectrum. CR systems try to improve the spectrum efficiency by using the spectrum holes in frequency, time, and space domains \cite{Zhao07asurvey,Akyildiz06survey}. This means that secondary users (SUs) are allowed to utilize the spectrum, provided that their transmissions do not interfere with the communication of primary users (PUs) \cite{Hykin05CR}. The fundamental components of CR systems that allow them to avoid interference are spectrum sensing and resource allocation. \par
However, in a practical CR network, spectrum occupancy measurements for all the frequency channels at all times are not available. This is partially because of energy limitations and network failures. Another highly important and very common reason for occurrence of missing entries in the data set is the hardware limitation. Each SU may want to use different frequency channels, but it may not be capable of sensing all the channels simultaneously \cite{Akyildiz08SS, Shah1312:Clustering}. On the other hand, a complete and reliable spectrum sensing data set is needed for a reliable resource allocation. Therefore, we need to develop a method to estimate the missing spectrum sensing measurements. This task is especially more challenging in dynamic environments.\par

There are different approaches toward the problem of data analysis in the CR networks. In \cite{Lambotharan13SVM}, a learning approach is introduced based on support vector machine (SVM) for spectrum sensing in multi-antenna cognitive radios. SVM classification techniques are applied to detect the presence of PUs.
Several algorithms have been been proposed using dictionary learning framework \cite{Joneidi14Outlier,GIANNAKIS13DICTIANARY}. These approaches try to find the principal components of data using dictionary learning  and exploit the components to extract information.\par

The goal of this paper is to estimate the missing spectrum sensing data as accurate as possible in the time varying environments. An approach is introduced based on Nonnegative Matrix Factorization (NMF) \cite{Paatero94PMF,NMFNATURE} to represent the spectrum measurements as additive, not subtractive, combination of several components. Each component reflects signature of one PU, therefore  the data can be factorized as the product of signatures matrix times an activation matrix. \par

Dimension reduction is an inevitable pre-processing step for high dimensional data analysis \cite{rahmani2015randomized}. NMF is a dimension reduction technique that has been employed in diverse fields \cite{Fevotte13SmoothNMF,hu2012robustNMF}. The most important feature of NMF, which makes it distinguished from other component analysis methods, is the non-negativity constraint. Thus  the original data can be represented as additive combination of its parts.\par 
In our proposed method, a new framework is introduced to decompose the spectrum measurements in CR networks using a piecewise constant NMF algorithm in presence of missing data. Piecewise constant NMF and its application in video structuring is introduced in \cite{Fevotte14Piecewise}. In the proposed method, we try to handle the missing entries in the data and also take a different approach to solve the corresponding optimization problem using an iterative reweighed technique.  \par

In the context of CR networks, NMF is utilized in \cite{Fu14Tensor} to estimate the power spectra of the sources in a CR network by factorizing the Fourier Transform of the correlation matrix of the received signals. Our proposed method estimates the missing entries in power spectral density measurements by enforcing a temporal structure on the activity of the PUs and can be used in scenarios when the number of the PUs is not known. \par

The introduced method takes advantage of a prior information about the activity of the PUs and exploits piecewise constant constraint to improve the performance of the factorization. Moreover, a solution for the introduced minimization problem is suggested using the Majorization-Minimization (MM) framework. \par
The rest of the paper is organized in the following order. In Section~\ref{model}, the system model and the problem statement are introduced. Section~\ref{PCNMF} describes the proposed new NMF problem. In Section~\ref{MM}, a method is presented to solve the piecewise constant NMF problem in MM framework with missing data. Section~\ref{results} presents the simulation results and finally Section~\ref{Conclusions} draws conclusions.\par
\section{System Model}\label{model}
Due to the nature of wireless environments,  trustworthy information cannot be extracted from measurements of a single SU. To find the spectrum holes in frequency, time, and space, there exists a fusion center that collects and combines the measurements from all the SUs \cite{Akyildiz08SS}. Cooperative spectrum sensing makes the missing data estimation algorithm more robust. Fusion center predicts the missing entries by using the collected measurements.
However, since each SU is not able to sense the whole spectrum all the time, the data set collected from the SUs contains missing entries. Network failures, energy limitations, and shadowing can also cause loss of data.\par

\begin{figure}[h]
\centering
\includegraphics[width=3.2in,angle=0]{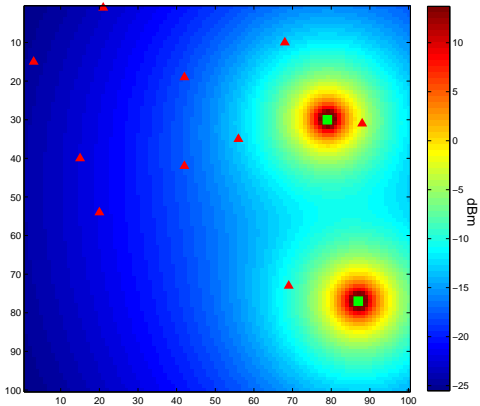}\vspace{-1.5mm}
\caption{\small{The power distribution of $2$ PUs (squares) and deployment of $10$ SUs (triangles) without considering shadowing effect.}}\label{powermap}
\end{figure}

Without loss of generality, we want to reconstruct the power map in a single frequency band. The network consists of $N_{P}$ primary transmitters and $N_{R}$ spectrum sensors that are randomly spread over the entire area of interest. Figure ~\ref{powermap} illustrates an example of a network with $N_{P} = 2$ PUs and $N_{R} = 10$ SUs in a $100 \times 100$ area.

The received power of the $r^{th}$ sensor at time $t$ can be written as
\begin{equation}\label{single_SU}
s_r(t) = \sum_{j = 1} ^{N_{P}} {p_j(t)\gamma_{rj}(t)} + z_r(t),
\end{equation}
where $p_j(t)$ is the transmit-power of the $j^{th}$ PU at time $t$, $\gamma_{rj}$ is the channel gain from the $j^{th}$ PU to the  $r^{th}$ SU, and $z_r(t)$ is the zero-mean Gaussian measurement noise at the $r^{th}$ sensor with variance $\sigma_r^2$. Considering a Rayleigh fading model, the channel gain coefficient can be modeled as:
\begin{equation}\label{channel_gain}
\gamma_{rj} = \frac{C{\lvert h_{rj} \rvert}^2}{{d_{rj}}^\alpha},
\end{equation}
where the channel constant $C = \frac{G_P G_R c^2}{(4 \pi f)^2}$, $f$ is the carrier frequency, $c$ is the speed of light, and $G_P$ and $G_R$ are the transmitter and receiver antenna gains. $\alpha$ is the path loss exponent which determines the rate at which power decays with the separation distance $d_{rj}$ between the $r^{th}$ SU and the $j^{th}$ PU and $ {\lvert h_{rj} \rvert}^2 $ models the fading effect. \par
At time slot $t$, measurements from SUs can be stacked in a vector $\boldsymbol{s}(t)$, given as
\begin{equation}\label{multiple_SU}
\boldsymbol{s}(t) = \sum_{j = 1} ^{N_{P}} {p_j(t)\boldsymbol{\gamma}_{j}(t)} + \boldsymbol{z}(t),
\end{equation}
where $\small{\boldsymbol{s}(t) = \left[ \begin{array}{cccc}
s_1(t) & s_2(t) & ... & s_{N_{R}}(t) \end{array} \right]^T}$, $\small{\boldsymbol{\gamma}_{j}(t) = \left[ \begin{array}{cccc}
\gamma_{1j}(t) & \gamma_{2j}(t) & ... & \gamma_{N_{R}j}(t) \end{array} \right]^T}$, and $\small{\boldsymbol{z}(t) = \left[ \begin{array}{cccc}
z_1(t) & z_2(t) & ... & z_{N_{R}}(t) \end{array} \right]^T}$. At each time slot, only a few SUs observe the power levels and report them to the fusion center. Therefore the vector $\boldsymbol{s}(t)$ contains some \emph{missing} entries. Furthermore, each PU can be active or inactive in each time slot.\par

Some of the characteristics of the environment can be exploited to simplify the problem. It is assumed that channel gains are slowly time varying such that they can be considered as constant in a time window. Therefore, matrix representation of (\ref{multiple_SU}) can be written as:
\begin{equation}\label{Compact_form}
\boldsymbol{S} =  \boldsymbol{\Gamma} \boldsymbol{P} + \boldsymbol{Z},
\end{equation}
where $\boldsymbol{S}$ is an $N_{R} \times T$ matrix, which includes measurements from sensors in $T$ time slots, $\boldsymbol{\Gamma}=[\boldsymbol{\gamma_{1}},\ldots,\boldsymbol{\gamma_{N_{P}}}]$ is a $N_{R} \times N_{P}$ matrix, which consists of $N_{P}$ channel gain vectors in the $N_{R}$-dimensional space of data, and $\boldsymbol{P}$ is an $N_{P} \times T$ matrix that indicates the power levels of PUs in each time slot ($p_{jt} = 0$  if the  $j^{th}$ PU is inactive at time $t$).\par
Here, the goal is to estimate the missing data using the partial observations. To achieve this goal, the data is decomposed using piecewise constant NMF. Then the components of data and the activation matrix are used to estimate the missing data.\par
\section{PC-NMF: Piecewise Constant Nonnegative Matrix Factorization}\label{PCNMF}
\begin{figure}[h]
\centering
\includegraphics[width=3.2in,angle=0]{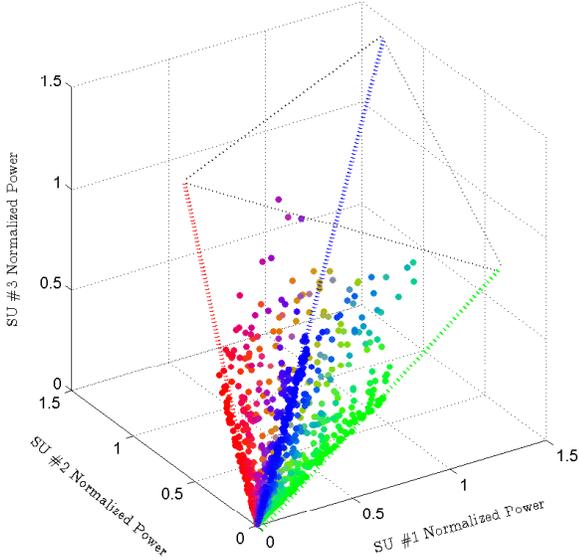}\vspace{-1.5mm}
\caption{\small{Structure of data generated by $N_{P} = 3$ PUs in $N_{R} = 3$ dimensional space. }}
\label{pyramid}
\end{figure}

Promoted by (\ref{Compact_form}), it is easy to see that the measurements of each time slot can be represented as an additive, not subtractive, combination of few vectors. This algebraic representation has a geometric interpretation. Figure ~\ref{pyramid} helps us to visualize the structure of data in a $3$-dimensional space of data. In this figure, $3$ SUs are measuring power levels in an area with $3$ PUs. It is easy to notice that measurement vectors lie within a pyramid in the positive orthant with $N_{P} = 3$ edges proportional to $\boldsymbol{\gamma_{j}}$. This is due to fact that all the points in the pyramid can be written as an additive combination of the edges. \par

Although it is assumed that the channel gains are stationary for a time window of length $T$, PUs can become activated/deactivated in this time window any number of times and can change their transmission power in each activation. Hence, the power levels of PUs tend to be piecewise constant. \par

NMF is a widely-used technique to decompose data to its nonnegative components. Here, the structure of the power level matrix $\boldsymbol{P}$ is exploited  while handling the missing entries. As a result, the general objective function is presented 
as follows:
\begin{equation}\label{general_objective}
\begin{aligned}
& \underset{\boldsymbol{\Gamma},\boldsymbol{P}}{\text{minimize}  }
& & { D_W (\boldsymbol{S} | \boldsymbol{\Gamma} \boldsymbol{P}) + \beta F(\boldsymbol{P}) },  \\
& \text{subject to}
& & \boldsymbol{\Gamma} \geq 0 , \boldsymbol{P} \geq 0,
\end{aligned}
\end{equation}
where $D_W (\boldsymbol{S} | \boldsymbol{\Gamma} \boldsymbol{P})$ is a weighted measure of fit and $F(\boldsymbol{P})$ is a penalty, which favors piecewise constant solutions. $\beta$ is a nonnegative scalar weighting the penalty. The constraints denote that all the entries of $\boldsymbol{\Gamma}$ and $\boldsymbol{P} $  are nonnegative. $W$ is an $N_{R} \times T$ weight matrix that is used to estimate the weighted distance between $\boldsymbol{S}$ and $\boldsymbol{\Gamma} \boldsymbol{P}$. The coefficients of the weight matrix denote the presence of data ($w_{rt} = 0 / w_{rt} = 1 $ if the measurement of the $r^{th}$ SU at time slot $t$ is unavailable/available).   \par

NMF algorithms utilize different measures of fit such as Euclidean distance, generalized Kullback-Leibler (KL) divergence, and the Itakura-Saito divergence . In all the cases, the distance can be calculated as the sum of the distances between different coefficients \cite{lee2001algorithms,fevotte2011algorithms,Fevotte13bayesian}.\par
\begin{equation}\label{seperable_measure}
D_W (\boldsymbol{S} | \boldsymbol{\Gamma} \boldsymbol{P}) = \sum_{t = 1} ^{T} {\sum_{r = 1} ^{N_{R}}} {w_{rt} d (s_{rt} | \sum_{j = 1} ^{N_{P}} {\gamma_{rj} p_{jt})}} ,
\end{equation}

In our case, Euclidean Distance is used as the measure of fit. This objective function is commonly used for problems with Gaussian noise model, a common noise model in communication systems, hence:
\begin{equation}\label{Euc_dist}
d (s_{rt}| \sum_{j = 1} ^{N_{P}} {\gamma_{rj} p_{jt}}) = \frac{1}{2}(s_{rt} - \sum_{j = 1} ^{N_{P}} {\gamma_{rj} p_{jt}})^2 .
\end{equation}

Since there exist sharp transitions in power level of PUs and power level of each PU is constant in each transmission period, rows of $P$ tend to be piecewise constant. In order to favor the piecewise constant solutions, the penalty function is defined as:
\begin{equation}\label{continuity_penalty}
F(\boldsymbol{P}) = \sum_{t = 2} ^{T} \sum_{j = 1} ^{N_{P}} { \lim _{n \rightarrow 0} | p_{jt} - {p}_{j(t-1)} | ^n }.
\end{equation}

When $n$ tends to $0$, this penalty function represents the sum of $\ell_0$ norm of the transition vectors, i.e. $\boldsymbol{p}_{t} - \boldsymbol{p}_{(t-1)}$, where
$\boldsymbol{p}_{t}$ is an $N_{P} \times 1$ vector containing power levels of PUs in time $t$.
This penalty favors the solutions with a lower number of transitions. However, since it is not differentiable, it can be replaced with a differentiable approximation:
\begin{equation}\label{continuity_penalty_epsilon}
\begin{aligned}
& F_\epsilon(\boldsymbol{P}) = \sum_{t = 2} ^{T} \sum_{j = 1} ^{N_{P}} { \rho_\epsilon( p_{jt} - {p}_{j(t-1)}) }, \\
& \text{with} \qquad \rho_\epsilon(x) = \frac{x^2}{x^2 + \epsilon^2},
\end{aligned}
\end{equation}
where $\epsilon^2$ is a small positive constant and is much less than all the non-zero elements of $(p_{jt} - p_{j(t-1)})^2$ $\forall j,t$ to avoid division by zero. 

In Section \ref{MM}, an algorithm is derived to find the minimizer of the following problem:
\begin{equation}\label{scale_objective}
\begin{aligned}
& \underset{\boldsymbol{\Gamma},\boldsymbol{P}}{\text{minimize}  }
& & { D_W (\boldsymbol{S} | \boldsymbol{\Gamma} \boldsymbol{P}) + \beta F_\epsilon(\boldsymbol{P}) },  \\
& \text{subject to}
& & \boldsymbol{\Gamma} \geq 0 , \boldsymbol{P} \geq 0. \\
\end{aligned}
\end{equation}

After estimating $\boldsymbol{P}$ and $\boldsymbol{\Gamma}$, the missing entries of $\boldsymbol{S}$ can be approximated using the equation  $\boldsymbol{\widehat{S}} \simeq \boldsymbol{\Gamma} \boldsymbol{P}$. \par

\section{Majorization-Minimization for Piecewise Constant NMF}\label{MM}
In this section, an iterative algorithm is described to find the solution of the optimization problem proposed in (\ref{scale_objective}). For that, Majorization-Minimization (MM) framework is employed \cite{lee2001algorithms,fevotte2011algorithms}. 
MM algorithm and its variants have been used in various applications such as parameter learning and image processing \cite{mm1,mm2}.
The update rules are derived to calculate the entries of $\boldsymbol{P}$ given the entries of $\boldsymbol{\Gamma}$ and then the entries of $\boldsymbol{\Gamma}$ given the entries of $\boldsymbol{P}$, using an  iterative reweighed algorithm.\par

First, the update rules for $\boldsymbol{P}$ given $\boldsymbol{\Gamma}$ are derived. Then, the update rules for $\boldsymbol{\Gamma}$ will be derived in a similar manner. \par 

As it is clear in (\ref{seperable_measure}), we can write the distance measure as a sum of different time slots:
\begin{equation}\label{C_p_t}
D_W (\boldsymbol{S} | \boldsymbol{\Gamma} \boldsymbol{P}) = \sum _{t = 1} ^{T} C(\boldsymbol{p}_t),
\end{equation}
where $C(\boldsymbol{p}_t)$ is the weighted Euclidean distance between $\boldsymbol{s}_t$ and $\boldsymbol{\Gamma}\boldsymbol{p}_t$, given $\boldsymbol{s}_t$ and $\boldsymbol{\Gamma}$. In MM framework, the update rules are derived by minimizing an auxiliary function \cite{lee2001algorithms}. By definition, $G(\boldsymbol{p}_t,\widehat{\boldsymbol{p}}_t)$ is an auxiliary function of $C(\boldsymbol{p}_t)$ if and only if $G(\boldsymbol{p}_t,\widehat{\boldsymbol{p}}_t) \geq C(\boldsymbol{p}_t)$ and $G(\boldsymbol{p}_t,\boldsymbol{p}_t) = C(\boldsymbol{p}_t)$ for $\forall \boldsymbol{p}_t$. If $G(\boldsymbol{p}_t,\widehat{\boldsymbol{p}}_t)$ is chosen such that it is easier to minimize, the optimization of $C(\boldsymbol{p}_t)$ can be replaced with iterative minimization of $G(\boldsymbol{p}_t,\widehat{\boldsymbol{p}}_t)$ over $\boldsymbol{p}_t$. Thus, in the literature, convex functions are frequently used as the auxiliary functions \cite{lee2001algorithms,Fevotte13SmoothNMF}. It is shown in \cite{lee2001algorithms} that $C(\boldsymbol{p}_t)$ is non-increasing under the update 
\begin{equation}\label{general_update_rule}
\begin{aligned}
&    \boldsymbol{p}_t^{i+1} = \underset{\boldsymbol{p}_t}{\text{argmin}}
& & G(\boldsymbol{p}_t,\boldsymbol{p}_t^{i}).
\end{aligned}
\end{equation}

This is due to the fact that in the $i^{th}$ iteration we have $C(\boldsymbol{p}_t^{i+1}) \leq G(\boldsymbol{p}_t^{i+1},\boldsymbol{p}_t^{i}) \leq G(\boldsymbol{p}_t^{i},\boldsymbol{p}_t^{i}) = C(\boldsymbol{p}_t^{i})$. \par

Following a similar approach as \cite{lee2001algorithms}, the auxiliary function for the weighted Euclidean distance can be formulated as: 
\begin{equation}\label{Aux_Euc}\small{}
    G(\boldsymbol{p}_t,\boldsymbol{p}_t^{i}) = C(\boldsymbol{p}_t^{i}) + (\boldsymbol{p}_t - \boldsymbol{p}_t^{i}) \nabla C(\boldsymbol{p}_t^{i}) + \frac{1}{2} (\boldsymbol{p}_t - \boldsymbol{p}_t^{i})^T K(\boldsymbol{p}_t^{i}) (\boldsymbol{p}_t - \boldsymbol{p}_t^{i}),
\end{equation}
where $K(\boldsymbol{p}_t^{i})$ is an $N_{P} \times N_{P}$ diagonal matrix with
\begin{equation}\label{Kjj}
\begin{aligned}
& k_{jj}(\boldsymbol{p}_t^i) = \frac{q_{jt}^i}{p_{jt}^i}, \\
& \boldsymbol{q}_t^i = \boldsymbol{\Gamma}^T(\boldsymbol{w}_t\odot\boldsymbol{\Gamma}\boldsymbol{p}_t^i),
\end{aligned}
\end{equation}
and $k_{jj}$ is the $j^{th}$ diagonal entry of $K(\boldsymbol{p_t^i})$ and $\odot$ is element-wise multiplication.\par 

To solve the problem presented in (\ref{scale_objective}), the contribution of $F_\epsilon(\boldsymbol{P})$ should be considered in the auxiliary function. For that, a convex version of $F_\epsilon(\boldsymbol{P})$ is employed:
\begin{equation}\label{reweighted_penalty}
\begin{aligned}
& F_\epsilon(\boldsymbol{P}) = \sum_{t = 2} ^{T} \sum_{j = 1} ^{N_{P}}  y_{jt}(p_{jt} - p_{j(t-1)})^2, \\
& \text{with} \qquad y_{jt} = \frac{1}{(p_{jt}^{(i-1)} - p_{j(t-1)}^{(i-1)})^2 + \epsilon}. 
\end{aligned}
\end{equation}

Now the update rules can be obtained using the iterative version of (\ref{reweighted_penalty}). This means that $y_{jt}$ is updated in each iteration using the values of $\boldsymbol{P}$ in the previous iteration. \par

To form the penalized auxiliary function,  $G_\beta(\boldsymbol{p}_t,\boldsymbol{p}_t^{i})$, we add up $G(\boldsymbol{p}_t,\boldsymbol{p}_t^{i})$ with the contribution of $\boldsymbol{p}_t$ to $F(\boldsymbol{P})$. Thus, $G_\beta(\boldsymbol{p}_t,\boldsymbol{p}_t^{i})$ can be written as:
\begin{multline}\label{G_beta}
G_\beta(\boldsymbol{p}_t,\boldsymbol{p}_t^{i}) = G(\boldsymbol{p}_t,\boldsymbol{p}_t^{i})   \\ 
+ \beta [\sum_{j = 1}^{N_{P}} y_{jt}^i(p_{jt} - p_{j(t-1)}^i)^2  \\ 
+ \sum_{j = 1}^{N_{P}} y_{j(t+1)}^i(p_{j(t+1)}^i - p_{jt})^2].
\end{multline}

It is worthwhile to mention that $y_{j1} = y_{j(T+1)} = 0$ $\forall j$. Since $G_\beta(\boldsymbol{p}_t,\boldsymbol{p}_t^{i})$ is convex, it can be easily minimized over $\boldsymbol{p}_t$ by setting the gradient to zero. Hence the update rule is attained as:
\begin{equation} \label{P_update_rule} \footnotesize
\begin{aligned}
& p_{jt}^{i+1} =  \frac{-\nabla_j C(\boldsymbol{p_t^i}) + p_{jt}^i k_{jj}(\boldsymbol{p}_t^i) + 2\beta y_{jt}^i p_{j(t-1)}^i + 2\beta y_{j(t+1)}^i p_{j(t+1)}^i }{k_{jj}(\boldsymbol{p}_t^i) + 2\beta y_{jt}^i  + 2\beta y_{j(t+1)}^i }, \\
& \nabla C(\boldsymbol{p_t^i}) = -\boldsymbol{\Gamma}^T(\boldsymbol{w_t}\odot\boldsymbol{s_t} - \boldsymbol{w_t}\odot(\boldsymbol{\Gamma}\boldsymbol{p_t^i}) ),
\end{aligned}
\end{equation}
where $\nabla_j C(\boldsymbol{p_t^i})$ is the $j^{th}$ element of the gradient $\nabla C(\boldsymbol{p_t^i})$.

Finding the update rule for $\boldsymbol{\Gamma}$ is simple. This is due to the fact that $F(\boldsymbol{P})$ is not a function of $\boldsymbol{\Gamma}$. Hence, the update rule for $\boldsymbol{\Gamma}$ is similar to the update rule for standard NMF, except the missing entries must be taken into account \cite{Mao04WNMF}. The update rules can be written in matrix form as:
\begin{equation} \label{Gamma_update_rule} \small{}
\boldsymbol{\Gamma}^{i+1} = \boldsymbol{\Gamma}^{i} \odot \frac{(\boldsymbol{W}\odot\boldsymbol{S})\boldsymbol{P}^T}{(\boldsymbol{W}\odot(\boldsymbol{\Gamma}^{i}\boldsymbol{P}))\boldsymbol{P}^T}
\end{equation}
where $\odot$ is the element-wise multiplication and the division is also performed in an element-wise manner. \par

The obtained update rules in (\ref{P_update_rule}) and (\ref{Gamma_update_rule}) are exploited alternatively to estimate $\boldsymbol{\Gamma}$ and $\boldsymbol{P}$. Then the missing entries of $\boldsymbol{S}$ are predicted by $\widehat{\boldsymbol{S}} = \boldsymbol{\Gamma}\boldsymbol{P}$. \par

However, by using the objective function in (\ref{scale_objective}), the optimization problem results in solutions with entries of $\boldsymbol{P}$ tend toward $0$ and $\|\boldsymbol{\Gamma}\|$ tends toward $\infty$.
We take advantage of the scale ambiguity between $\boldsymbol{\Gamma}$ and $\boldsymbol{P}$ to avoid this issue. 
Let $\boldsymbol{\Lambda}$ be a diagonal $N_{P} \times N_{P}$ matrix with its $j^{th}$ diagonal entry equal to $\| \boldsymbol{\gamma}_j \|_2$. In each iteration, the rescaled matrix pair $(\boldsymbol{\Gamma}\boldsymbol{\Lambda}^{-1},\boldsymbol{\Lambda}\boldsymbol{P})$ is used instead of the original matrix pair $(\boldsymbol{\Gamma},\boldsymbol{P})$. \par 

As a practical scenario, we should also consider the case when the secondary network has no information about the number of the PUs, i.e. $N_{P}$. In this case, the common dimension of matrices $\boldsymbol{\Gamma}$ and $\boldsymbol{P}$ is not known. There have been some efforts in model order selection in NMF \cite{Fevotte13bayesian}. In the numerical experiments, $K > N_{P}$ is used as the common dimension to factorize the data in such conditions. This is only possible if the secondary network has some information about the upper bound of $N_{P}$.\par


\section{Numerical Results}\label{results}
\begin{figure}[h]
\centering
\includegraphics[width=3in,angle=0]{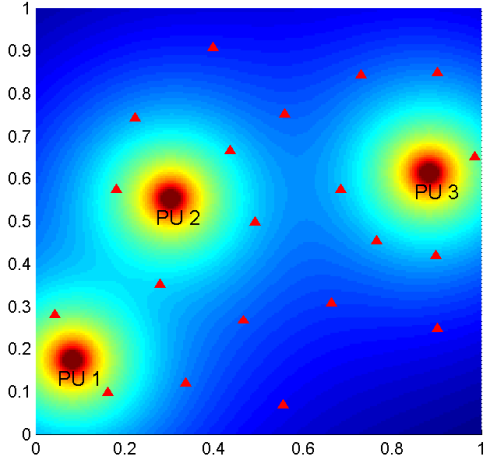}\vspace{-1.5mm}
\caption{\small{Network topology consisting of 3 PUs and $N_{R} = 20$ SUs marked by triangles. }}
\label{topol}
\end{figure}

For the numerical experiments, one frequency channel is considered with $N_{p} = 3$ active PUs in the area. Figure \ref{topol} illustrates the topology of the network. Incomplete measurements are collected from $N_{R} = 20$ SUs. \par

We use the same simulation environment and the same network topology as in \cite{GIANNAKIS13DICTIANARY}. The simulation parameters are set as follows, unless otherwise is stated. The path loss is computed as $(\frac{d}{d_0})^\alpha$, where $d$ is the distance, $d_0 = 0.01$,  and $\alpha = 2.5 $. $\gamma_{rj}$ is computed by multiplying the pathloss by the fading coefficient $\lvert h_{rj} \lvert ^2$ where\par
\begin{equation}
h_{rj}(t) = \eta h_{rj}(t-1) + \sqrt{1-\eta^2}\nu_{rj}(t),
\end{equation}
$\eta = 0.9995$, and $\nu_{rj}(t)$ is circulary symmetric zero mean complex Gaussian noise with variance $1$ \cite{GIANNAKIS13DICTIANARY}.

PUs' activity is modeled by a first order Markov model. All the PUs utilize the spectrum $\lambda = 0.3$ of the time slots. Transition matrix of the $j^{th}$ PU is $ \small \left( \begin{array}{cc} 1-a_j & a_j \\ b_j & 1-b_j \end{array} \right) $ and $\lambda = \frac{b_j}{a_j + b_j}$. $a_j$ is the probability that the $j^{th}$ PU stops transmitting from time $t-1$ to $t$ and $b_j$ is the probability that the PU activates transmitting. The parameter $a_j$ is uniformly distributed over $[0.05,0.15]$. \par

Each time a PU becomes activated, it chooses the transmission power from a uniform distribution with support $[100,200]$. Each SU makes a measurement with $70\%$ of chance. The measurements are contaminated by additive white Gaussian noise. The noise variance is $10^{-5}$  for all the SUs.\par

Partial measurements are generated for $ T = 600$ time slots. To reduce the computational burden, the first $300$ time slots are used to estimate $\boldsymbol{\Gamma}$. Next, by using the obtained $\boldsymbol{\Gamma}$ and the update rule (\ref{P_update_rule}), $\boldsymbol{P}$ is estimated for all $600$ time slots. The regularization factor $\beta$ is set to $5 \times 10^{-3}$ and $K = 5$ factors are used to factorize the data.\par

\begin{figure}[h!]
\centering
\includegraphics[width=3.4in,angle=0]{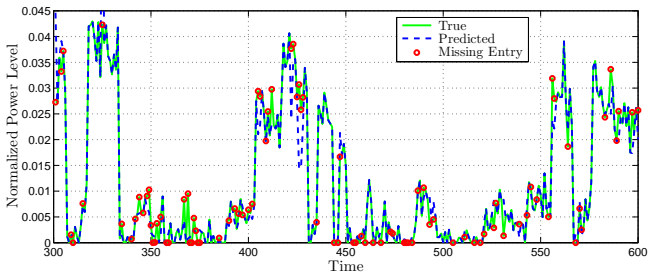}\vspace{-1.5mm}
\caption{\small{Power levels of a single SU.}}
\label{TrueVSPredicted}
\vspace{-2mm}
\end{figure}

Figure \ref{TrueVSPredicted} shows the true power levels and the reconstructed one at a randomly selected SU versus time for the time window of $T=300$ samples. It can be seen that the missing entries are accurately recovered through the proposed method, and it is evident that the proposed algorithm can easily track abrupt changes in power level. \par 

\begin{figure}[h]
    \centering
    \begin{subfigure}[b]{1\columnwidth}
        \centering
        \includegraphics[width=3.4in,angle=0]{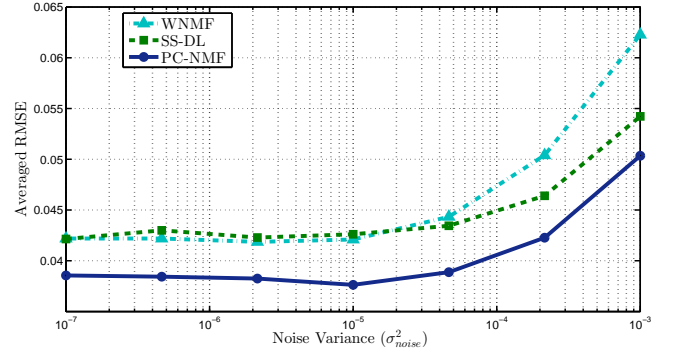}
        \label{vsNoise}
        \caption{\small $P_{miss} = 0.3$ , $K = 5$}
    \end{subfigure}
    \vfill
    \vspace{2.5 mm}
    \begin{subfigure}[b]{1\columnwidth}
        \centering
        \includegraphics[width=3.4in,angle=0]{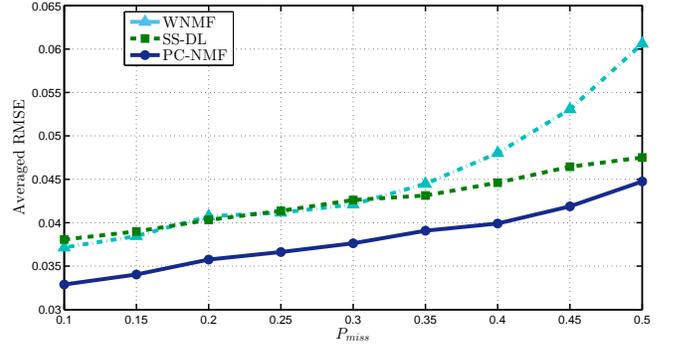}
        \label{vsPmiss}
        \caption{\small $\sigma_{noise}^2 = 10^{-5}$ , $K = 5$}
     \end{subfigure}    
    \caption{\small Performance of the proposed method for different noise levels and probability of miss, averaged over $200$ Monte Carlo trials. }
    \label{NMF_DL}
    \vspace{-2mm}
\end{figure}

Figure \ref{NMF_DL} compares the RMSE, averaged over SUs, of the proposed method with two similar methods. The method introduced in \cite{GIANNAKIS13DICTIANARY} exploits the spatial correlation between adjacent SUs' measurements and semi-supervised dictionary learning (SS-DL) to estimate the missing entries. For the numerical results, the batch version of SS-DL is employed and the parameters are set to their optimal values. Furthermore, to emphasize the effect of the piecewise constant penalty, the results are also compared with the weighted NMF, i.e. WNMF \cite{Mao04WNMF,Kim09WNMF}. WNMF employs binary weight matrix to deal with the missing entries. This figure shows that the proposed method outperforms its competitors in different noise levels (Figure \ref{NMF_DL}.(a)) and different probabilities of miss (Figure \ref{NMF_DL}.(b)). $P_{miss}$ denotes the ratio of the missing entries among the spectrum data.\par

This figure shows that WNMF and SS-DL almost perform the same for low noise variance and low $P_{miss}$. However, for harsh environments with high noise variance or high $P_{miss}$, SS-DL produces more accurate results. The PC-NMF method outperforms both methods in different noise levels and different probabilities of miss. For instance, PC-NMF has $11.6 \%$ less RMSE compared to the SS-DL method for $\sigma_{noise}^2 = 10^{-5}$ and $P_{miss} = 0.3$.

This improvement in the performance does not increase the computational burden of the algorithm. Table \ref{TIME} shows the running times\footnote{All simulations have been performed under MATLAB 2014a environment on a PC equipped with Intel Xeon E5-1650 processor (3.20 GHz) and 8 GB of RAM.} for different methods averaged over $100$ Monte Carlo trials for $\sigma_{noise}^2 = 10^{-5}$ , $P_{miss} = 0.3$, and $K = 5$. 

\begin{table}[h!]
    \caption{\small{Average Running Time}}
    \begin{center}
        \begin{tabular}{| c | c |}
        \hline 
        \textbf{Method} & \textbf{Average Running Time (s)} \\ \hline
        SS-DL & $10.3039$  \\ \hline
        WNMF & $0.0944$ \\ \hline
        PC-NMF & $0.0952$ \\ \hline
        \end{tabular}
    \label{TIME}    
    \end{center}
    \vspace{-2mm}
\end{table}

 It is known that the NMF methods converge much faster than methods based on gradient descent \cite{Mao04WNMF}. However, Table \ref{TIME} also illustrates that the proposed method does not require more computational resources compared with WNMF. \par   
 
\begin{figure}[h!]
\centering
\includegraphics[width=3.4in,angle=0]{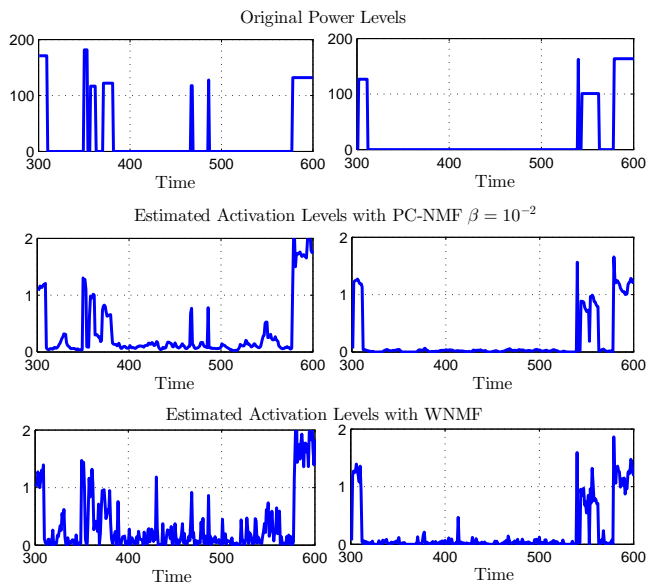}\vspace{-1.5mm}
\caption{\small{Original power levels of different PUs and the estimated activation levels with PC-NMF and WNMF.}}
\label{PUpowerlevel}
\vspace{-2mm}
\end{figure}

To study the effect of the piecewise constant penalty on the output of the algorithm, Figure \ref{PUpowerlevel} depicts the power level of two PUs and the estimated activation levels using the introduced method and WNMF. Both methods can estimate the power levels up to a scale factor. The number of factors is set to $3$, i.e. $K = N_{P}$.\par 

This figure illustrates the fact that the proposed method produces a more accurate factorization by taking advantage of piecewise constant constraint as a prior information. As it was expected, power levels estimated by PC-NMF are piecewise constant, while the results generated by WNMF are noisy. In fact, the piecewise constant penalty decreases the effect of noise and fading. Moreover, the sharp transitions are preserved in the factors returned by PC-NMF.\par
\vspace{-2mm}

\section{Conclusions}\label{Conclusions}
By exploiting inherent structural feature of cognitive radio networks, we proposed a piecewise constant NMF approach that can decompose the data set into its components. Majorization-Minimization framework is utilized to solve the optimization problem of the piecewise constant NMF. Numerical simulations suggest that this method is able to predict the missing entries in the spectrum sensing database accurately.\par

\footnotesize
{
\bibliography {NMF}
\bibliographystyle{ieeetr}
}
\end{document}